\documentclass[twocolumn,aps]{revtex4}
\usepackage{epsfig}
\usepackage{psfig}
\usepackage{amsmath}

\newcommand{\be}{\begin{equation}}   
\newcommand{\ee}{\end{equation}}   
\newcommand{\bea}{\begin{eqnarray}}   
\newcommand{\eea}{\end{eqnarray}}

\begin{document}
\title{Is the Universe odd?}
\author{Kate Land and Jo\~{a}o Magueijo }
\address{Theoretical Physics Group, Imperial College, Prince Consort Road, 
London SW7 2BZ, UK}

\begin{abstract}
We investigate the point-parity and mirror-parity handedness of the large
angle anisotropy in the cosmic microwave background (CMB). In particular
we consider whether the observed low CMB quadrupole could more generally signal
odd point-parity, i.e. suppression of even multipoles.
Even though this feature is ``visually'' present in most renditions
of the WMAP dataset we find that it never supports parity preference 
beyond the meagre $95\%$ confidence level. This is fortunate as 
point parity handedness implies almost certainly a high level of 
galactic contamination.
Mirror reflection  parity, on the contrary, is related to the emergence 
of a preferred axis, defining the symmetry plane. We use this technique
to  make contact with recent claims for an anisotropic Universe,
showing that the detected preferred axis is associated with positive (even) 
mirror parity. This feature may be an important clue in identifying the
culprit for this unexpected signal.
\end{abstract}

\pacs{PACS Numbers: *** }
\date{\today}

\maketitle


The properties of physical systems when subject to parity transformations 
are of great interest and the concept has been extensively used in chemistry,
particle physics and condensed matter systems. In this paper we examine 
the parity properties of the large angle cosmic microwave background
(CMB) temperature as rendered by the Wilkinson Microwave Anisotropy
Probe (WMAP)~\cite{wmap}. With the exception of the $S$ statistic proposed 
in~\cite{teg} this issue has strangely received almost no attention
(see also the theoretical work of~\cite{kam,lev}).
And yet there are a number of practical and theoretical considerations
that make this type of analysis very topical. One should consider 
separately two types of parity 
transformations: mirror reflections (i.e.: through a plane) and point 
reflections (relating antipodal points in the sky). They have 
very different implications.

Several anomalous features in the WMAP data have been 
reported~\cite{teg,copilet,ral,erik1,hbg,us,sour,erik2,erik3,hansen,vielva,evil},
pointing toward a preferred direction in the sky, 
the so-called ``axis of evil''.  The origin of this effect remains
mysterious, and it could well be that it is due to foreground
contamination or unsubtracted systematic errors.
Unlike point reflections, mirror reflections select a preferred 
direction in the sky, that of the normal to the symmetry plane.
Hence the search for mirror handedness entails the search for
a preferred axis in the CMB fluctuations (although the converse 
need not be true). 
The first purpose of this paper is to investigate whether mirror
parity could shed light upon the observed statistical anisotropy
of CMB fluctuations.

Should there be a preferred parity or handedness associated with 
the ``axis of evil''
effect such a fact could be crucial in identifying the culprit. It has
been suggested that the preferred axis could be the signature of
a non-trivial cosmic topology~\cite{riaz,sourtop,teg,dodec,sperg,evil},
anisotropic expansion~\cite{arj,tesse}, or even intrinsic cosmic 
inhomogeneity~\cite{moffatinh}. In this context the handedness of the sky, 
if  present, could supply selections for viable theoretical models.

A major open question in all of these studies is galactic foreground
contamination. This has been ruled out in a variety of ways but
remains an ongoing concern in all CMB work. Point reflection symmetry, 
in contrast with mirror symmetry, tests statistical homogeneity. 
Handedness with respect to point reflections could only be seen from
a cosmic vantage point, the focal point of the symmetry.
Such an observation would be cataclysmic for any theory of the Universe: 
even though  some topological and inhomogeneous models~\cite{moffatinh}
do violate translational invariance, that we might live in
a privileged point remains extremely unlikely in {\it any} cosmology.
But this leads to a very practical tool: if evidence for point 
reflection handedness were found this would most probably indicate
foreground galactic contamination, since with respect to
these the foregrounds we are in fact  ``at the centre
of the world''. Worryingly, the well documented low quadrupole 
(see e.g.\cite{bridle,contaldi,slosar}) could
be the tip of the iceberg revealing a preference for odd point parity,
and thus galactic contamination.
We investigate this matter quantitatively in this paper.

\begin{figure}
\centerline{\psfig{file=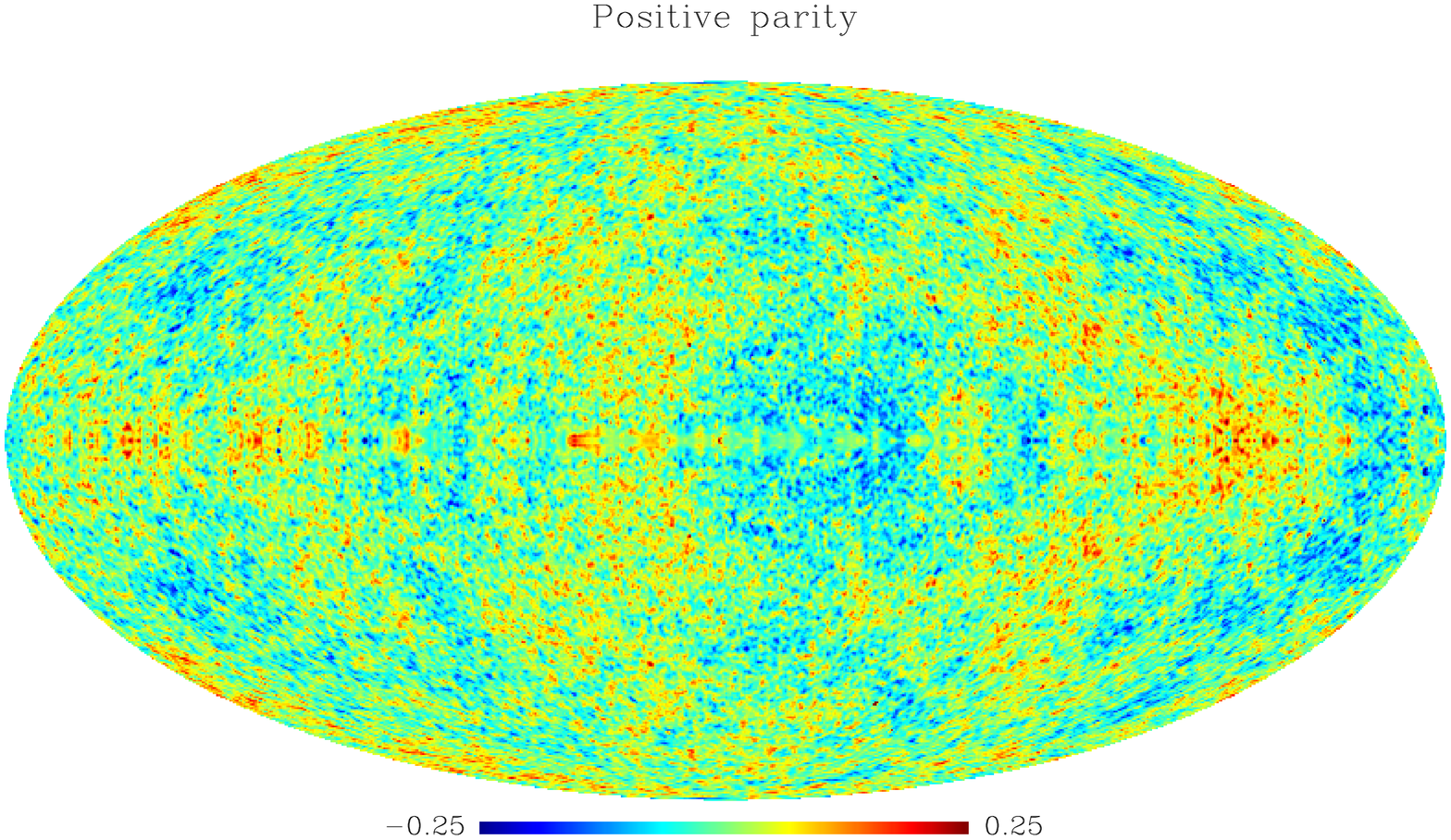,angle=90,width=9cm}}
\centerline{\psfig{file=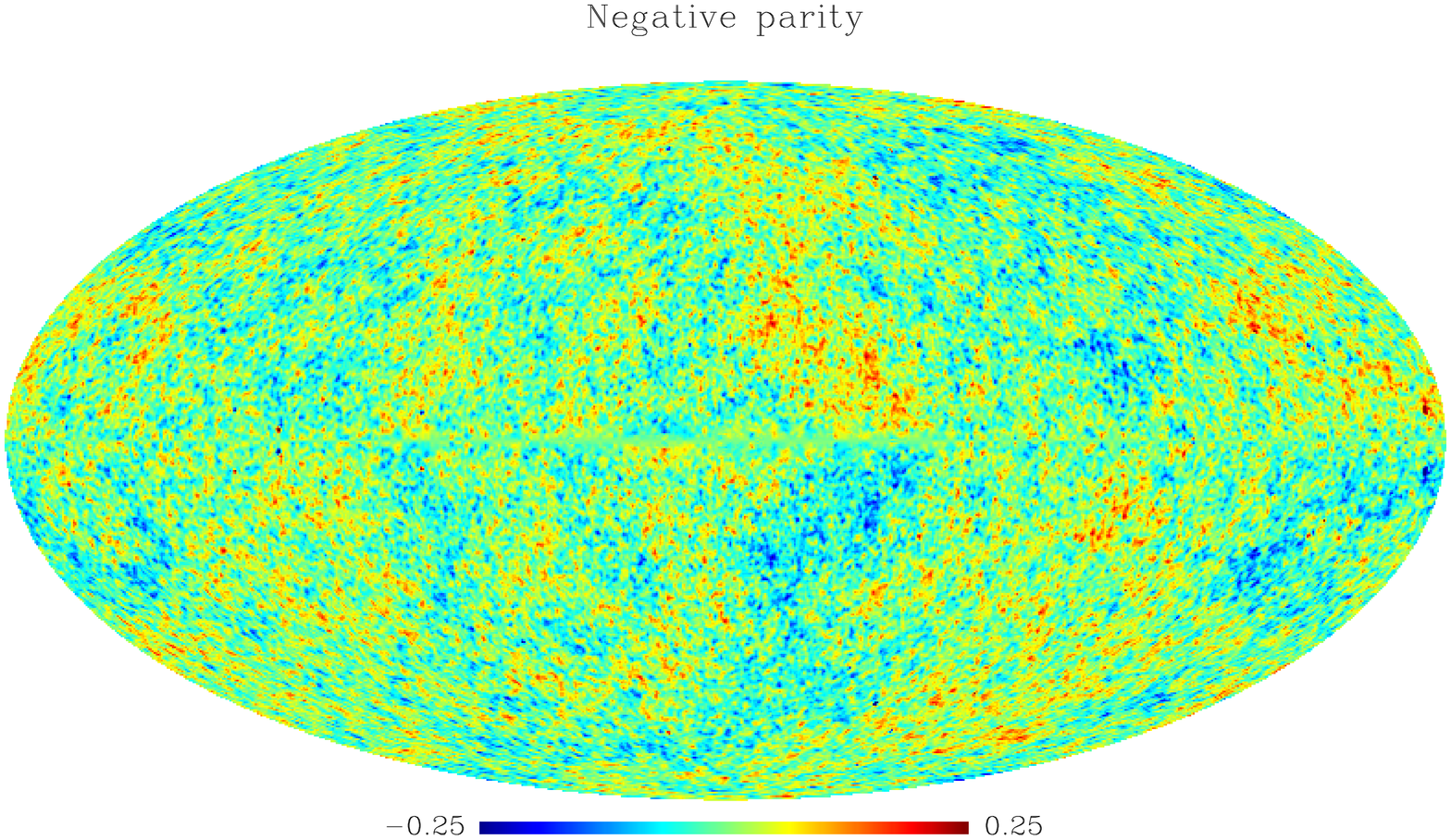,angle=90,width=9cm}}
\caption{The maps $M^+$ and $M^-$  for mirror transformations
with ${\mathbf n}$ aligned with the galactic poles (as derived 
from a foreground cleaned map).
}
\label{fig2}
\end{figure}

We define parity with respect to reflections through the origin 
as ${\mathbf x}'=-{\mathbf x}$ and 
for reflections through a plane as 
${\mathbf x}'={\mathbf x}-2({\mathbf x}\cdot {\mathbf n}){\mathbf n}$
(for a mirror with normal ${\mathbf n}$).
Let $P$ be one of these parity transformations. 
Then from a map $M$ one may extract a parity reversed map ${\tilde M}=PM$,
and define the positive and negative parity components:
\be
M^\pm={M\pm{\tilde M}\over 2}
\ee
One has that $M=M^++M^-$ and $PM^\pm=\pm M^\pm$.
As an illustration in Fig.~\ref{fig2} we take a rendition~\cite{teg0}
of the WMAP first year data and plot $M^+$ and $M^-$ for a mirror 
transformation with ${\mathbf n}$ aligned with the galactic pole.
A measure of handedness is generally a comparison between the 
two components $M^+$ and $M^-$.

\begin{figure}
\centerline{\psfig{file=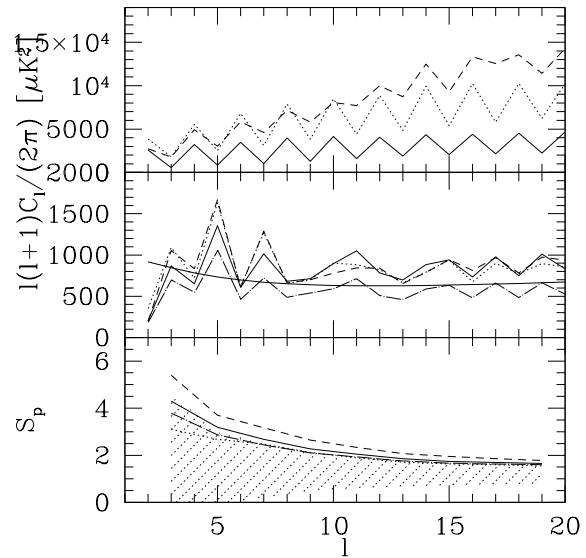,width=8cm}}
\caption{
Top panel: the power spectrum in the V band of 
templates for dust (solid), synchrotron (dotted) and
free-free (dash). Middle panel:
the measured power spectrum ${\hat C}_\ell$ of WMAP, against 
the best fit model (solid). We depict four renditions:
ILC (dashed), TOH cleaned (solid jagged), TOH Wiener filtered (dot-dash)
and LILC (dotted); see text for an explanation. Bottom panel:
the values of the $S_p$ statistics for the same datasets. The shaded
band is the ``1-sigma''.}
\label{figcl}
\end{figure}

For this purpose
we expand $M^+$ and $M^-$ into spherical harmonics, defined by:
\begin{eqnarray}
\frac{\Delta T}{T}(\hat{\mathbf r})=\sum_\ell \delta T_\ell=
\sum_{\ell m}a_{\ell m}Y_{\ell m}(\hat{\mathbf r})
\label{almdef}
\end{eqnarray}
and evaluate the power spectrum ${\hat C}_\ell$, defined as
$(2\ell +1){\hat C}_{\ell }=\sum_m |a_{\ell m}|^2$, for each of these
maps.
For point reflections only even (odd) $\ell$ multipoles appear in 
$M^+$ ($M^-$) and hence only their even (odd) ${\hat C}_\ell$ components 
are non-zero. A sign of point-parity handedness would therefore
be intermittency in the power spectrum, i.e. fluctuations
in power preferring alternate multipoles. Such a phenomenon would
enhance one of the maps $M^\pm$ with respect to the other. 

For mirror symmetries, on the other hand, only modes with even (odd) $\ell+m$
appear in $M^+$ ($M^-$). The power spectrum of $M^\pm$ is therefore
\be\label{clpmeq}
{\hat C}_\ell^\pm({\mathbf n_\ell})
={1 \over 2\ell +1}{\sum_{m}} p^\pm_{\ell m} |a_{\ell m}|^2
\ee
where $p^\pm_{\ell m}$ is 1 or 0 depending on the parity of
$\ell +m$. Here ${\mathbf n}_\ell$ is the $z$-axis
used to evaluate the expansion (\ref{almdef}), and we stress that
the decomposition into even and odd modes depends on its choice
(notice that ${\mathbf n}_\ell$ may be different 
for different $\ell$).
Mirror handedness is signalled
by a dominance of ${\hat C}_\ell^+$ over ${\hat C}_\ell^-$ or vice versa, 
and so it is another way of assessing whether some $m$ modes are preferred
over others (see~\cite{evil,conf,multipole}). 
As explained in~\cite{evil,conf,multipole} measures 
of $m$-preference are intrinsically measures of statistical anisotropy, 
since they are linked to the choice of ${\mathbf n}_\ell$.  Later on in this
paper we shall relate our proposal for a measure of asymmetry between
${\hat C}_\ell^+$ and ${\hat C}_\ell^-$ to other choices of measures of statistical
anisotropy and their choices of ${\mathbf n}_\ell$.

We now make concrete proposals for handedness statistics, starting with
point-parity. As the 
middle panel of Fig.~\ref{figcl}  shows, the power spectrum
$\ell(\ell+1){\hat C}_\ell$ displays a distinctive pattern of alternate 
low and high values,
starting with the much publicised low 
quadrupole~\cite{bridle,contaldi,slosar} and extending up to $\ell=9$.
Worryingly this is present in all renditions of the data.
As visually striking as this feature may be, it is important 
to quantify its significance. A possible statistic for intermittency
is:
\be
S_p=\sum_3^{l_{max}} {\ell(\ell+1){\hat C}_\ell\over \ell(\ell-1){\hat C}_{\ell-1}}
\ee
where the sum is over odd $\ell$ (i.e. considering ratios of
adjoining pairs, without overlap). $S_p$ measures point-parity preference 
for quasi-scale-invariant spectra, with $S_p\gg 1$ representing odd parity
and $S_p\ll 1$ even parity. The fact that the best-fit spectrum
is not scale-invariant induces a bias in $S_p$, but this is also
present in Monte-Carlo simulations performed to evaluate the significance
of any anomaly. 

As argued above point-parity preference would signal that
we live, as it were, in the centre of the world, a fact most easily
explained by foreground contamination. We now explain how point-parity 
may be used as a practical tool for detecting foregrounds.
In the top panel of Fig.~\ref{figcl}  we see that  galactic templates
do display a clear intermittency in power, favouring
even multipoles. This is promptly formalised by our statistic $S_p$:
for $\ell_{max}=19$ we find $S_p=0.37, 0.51, 0.80$ for dust,
synchrotron and free-free emissions respectively.  We have used the  
V band for definiteness, but other bands and choices of $\ell_{max}$ reveal 
the same strong signature of even-handedness. Apart from the free-free
map, these values are anomalous well beyond the 99\% confidence level.

\begin{table}
\begin{tabular}{lllll}
\hline
$\ell_{max}$  & $S_p$ & ${\overline S}_p$ & $\sigma (S_p)$& P(reject) \\
\hline
3&4.30&1.72&2.93&0.935\\
5&3.19&1.45&1.52&0.943\\
9&2.27&1.29&0.78&0.955\\
13& 1.85&1.22&0.53&0.948\\
19& 1.66&1.17&0.36&0.968\\
21& 1.57&1.16&0.32& 0.952\\
31&1.38&1.12&0.22&0.941\\
51&1.22&1.09&0.13&0.912\\
\end{tabular}
\caption{The observed value of the $S_p$ statistic in the 
TOHc map for different values of $\ell_{max}$, against its 
average value and variance as obtained from simulations, and confidence 
levels for detecting preferred odd point parity in the CMB. }
\label{tablepp}
\end{table}

In contrast, as already visually guessed, the CMB prefers 
odd point-parity, and Fig.~\ref{figcl} (bottom panel) reveals 
$S_p\gg 1$ for several $\ell$ ranges. If this feature were 
statistically significant it 
could be due to over-correcting for galactic foregrounds.
Fortunately this is not the case, as shown by Monte Carlo simulations.
To build intuition 
in Table~\ref{tablepp} we have displayed the average value of $S_p$ 
and its r.m.s. inferred from simulations of Gaussian maps
with the best fit power spectrum, subject to the WMAP noise
and beam. We then considered, as an example, the cleaned maps of~\cite{teg0} 
(which we denote TOHc) and evaluated $S_p$ for this map for a variety 
of $\ell_{max}$. By asking what percentage of simulations display
a larger $S_p$ we can evaluate the confidence level for detecting
even point-parity preference. This is never above 97\% and is in fact
below 95\% for the most visually striking features.
Thus what by eye appears as a very 
striking feature, is actually not significant under closer scrutiny.

We have considered other renditions of the dataset: 
the Wiener filtered maps of~\cite{teg0} (TOHw); the internal linear 
combination maps of~\cite{foregs} (ILC), and the Lagrange multipliers
internal linear  combination maps of~\cite{erik3}  (LILC).
In the bottom panel of Fig.~\ref{figcl} we plot $S_p$ for various
$l_{max}$ for all these maps, as well as the 1 sigma contour (which
should be interpreted with care given the skewed nature of the distribution).
We see that all maps have roughly consistent profiles with the exception
of the ILC map, which reveals systematically larger
values of $S_p$. This suggests that the ILC
map may be more contaminated by residual artifacts than other full-sky
renditions; still we never see evidence for odd point-parity
beyond the 97\% confidence level. 

\begin{figure}
\centerline{\psfig{file=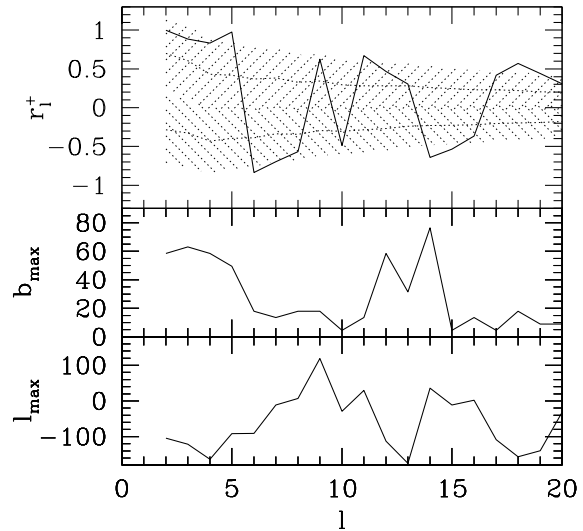,width=8cm}}
\caption{The statistic $r^+_\ell$ for the WMAP data and its distribution
inferred from Monte-Carlo simulations. This is bimodal so we plot the 
peak location and width for the 2 modes (corresponding to even and odd
realizations). The bottom two pannels supply the prefered axis 
${\mathbf n}_\ell$ in galactic coordinates $(b,l)$.}
\label{fig3}
\end{figure}

Reassured by this result we turn to mirror reflections. These 
depend on the choice of $z$-axis, and so are associated with or complementary
to any statistics seeking statistical anisotropy. In~\cite{evil} we 
advocated the use of
\be\label{rlm}
r_\ell=\max_{m{\bf n}} \frac{C_{\ell m}}
{(2\ell +1){\hat C}_\ell}
\ee
where $C_{\ell 0}=|a_{\ell 0}|^2$, $C_{\ell m}=2|a_{\ell m}|^2$ 
for $m>0$ (notice that 2 modes contribute for $m\neq 0$) .
This statistic provides three basic quantities:
the direction ${\mathbf n}_\ell$, the ``shape'' 
$m_\ell$, and the ratio $r_\ell$
of power absorbed by multipole $m_\ell$ in direction ${\bf n}_\ell$.
Essentially it seeks the direction ${\mathbf n}_\ell$ in which 
the highest ratio of power $r_\ell$ is concentrated in a 
single $m$-mode. Thus it is a statistic for both anisotropy
and $m$-preference.

We can use this statistic to select the direction $\mathbf n_\ell$
in which to evaluate ${\hat C}_\ell^\pm$  and  assess mirror
handedness. The asymmetry between odd and even modes  may then be measured 
by the ratio:
\be
r_\ell^+= {{\hat C}_\ell^+(\mathbf n_\ell)-{\hat C}_\ell^-(\mathbf n_\ell)
\over {\hat C}_\ell}
\ee
with ${\hat C}_\ell^\pm$ defined in (\ref{clpmeq}), and $\mathbf n_\ell$
defined by (\ref{rlm}).
This complements the work of~\cite{evil} in that it assesses whether 
or not an {\it existing} preferred axis is endowed with mirror parity
handedness. 
In~\cite{evil} we found that multipoles $\ell=2,...,5$ 
share a preferred axis, located roughly at $(b,l)\approx (60,-100)$
in galactic coordinates.
This extended earlier claims by~\cite{teg,copilet}, who noted
that $\ell=2,3$ are uncannily planar (i.e. $m=\pm\ell$ modes)
along this axis. We pointed out~\cite{evil} 
that the alignment of the preferred axes 
extends up to $\ell=5$ but the preferred ``shape'' is not
planar for $\ell=4,5$. The significance of preferred axes'
alignment is  at the 99.9\% level, when the problem is reanalized
from this perspective.

As Fig.~\ref{fig3} shows we may now add to this result  
the information that {\it all} aligned multipoles have even mirror-parity,
that is $r^+_\ell>0$. Even though the chisquared 
associated with these $r^+_\ell$ is not anomalous it is interesting to
notice that the observed $r^+_\ell$ are all above the average 
$r^+_\ell$ instead of scattering below and above it. 
We have evaluated
the distribution of $r_\ell^+$ from 5000
Monte Carlo simulations for Gaussian maps
with the best fit power spectrum, subject to the WMAP noise
and beam. This distribution is bimodal, i.e. there are two peaks one for
$r_\ell^+>0$, another for $r_\ell^+<0$. We therefore represented
two sets of ``average and error bars'' in  Fig.~\ref{fig3},
corresponding to these two peaks.

We can now ask what is the probability for the observed $r_\ell^+>0$,
in the range $\ell=2,...,5$ of aligned multipoles.
We find that on its own the observed handedness is not anomalous: 
indeed 10\% of the simulations reveal features as extreme as the
one observed. However this parity feature is found {\it in connection} 
with the  alignment of ${\mathbf n}_\ell$, which is indeed anomalous at
the 99.9\% significance level. It should therefore be regarded on 
the same footing as the alignment of
the phases in $a_{\ell m}$ reported in~\cite{evil}, which is not
unusual by itself (it's anomalous at the 97\% confidence level) 
but does become interesting in that
it qualifies the very anomalous alignment of the axes. 
We believe that the positive mirror parity reported in this paper
may be essential in identifying the
theoretical explanation for this effect.

In summary we have investigated the parity properties of the CMB
temperature anisotropy, distinguishing between point and mirror
parity. Point-parity handedness would almost
certainly be due to galactic foregrounds, thereby 
providing a contamination detection tool. We do detect even handedness
in the galactic templates, but our work was motivated by the 
the pattern of low-high values in the large angle ${\hat C}_\ell$,
pointing toward odd parity.
This might signal 
over-correcting for galactic emissions. 
Fortunately we don't find any evidence for odd parity
in publicly available full-sky maps once we
study the effect more quantitatively. Interestingly,
the ILC maps~\cite{foregs} have the strongest odd parity signal. 

Mirror parity was used as a complement to tests of statistical
isotropy. If the fluctuations select a preferred axis, as has been claimed,
we may ask if they also reveal mirror parity handedness. The answer
is yes: it appears that the ``axis of evil'' effect is endowed
with even mirror parity. Thus the planarity of the quadrupole and octupole
(corresponding to $\ell=2=|m|$ and $\ell=3=|m|$) is just 
an example for preferred even $\ell +m$ modes for $\ell=2,...,5$.
This is an interesting remark that may help in explaining the
observed result, should it not be due to unmodelled residual 
foregrounds or systematic errors. 

For example it has been suggested that a non-trivial topology
induces a large wave in the sky, $\Phi({\mathbf k,\eta_{ls}})$,
with wavelength just outside the horizon. This induces a
$m=0$ mode
$a_{\ell m}=A\sum\Phi({\mathbf k,\eta_{ls}})
i^\ell j_\ell(k\Delta\eta_{ls})Y^\star_{\ell m}
({\hat{\mathbf k}})$
which could be behind the observed axis (by destructive interference
with other modes). The parity of the axis imposes strong constraints
on the phase of this wave.
We are currently investigating this and other~\cite{tesse} possibilities, 
in the light of the findings on
mirror parity reported in this paper.

{\bf Acknowledgements} We thank C. Contaldi, K. Gorski, 
A. Jaffe and J. Medeiros 
for helpful comments. Some of the results in this paper 
were derived using the HEALPix package (\cite{healp}), and calculations
were performed on COSMOS, the UK cosmology supercomputer.

\label{lastpage}


\begin{thebibliography}{99}
\bibitem{wmap} Bennett C.L. et al., 2003, Astrophys. J. Suppl, 148, 1
\bibitem{teg} A. de Oliveira-Costa et al., 2004, Phys. Rev, D69, 063516
\bibitem{kam}A. Lue, L. Wang, M. Kamionkowski, Phys. Rev. Lett. 83:
1506-1509, 1999.
\bibitem{lev}L. Pogosian, T. Vachaspati, S. Winitzki,
New Astron.Rev. 47: 859-862, 2003.
\bibitem{copilet} Schwarz D. et al.,  Phys.Rev.Lett. 93: 221301, 2004.
\bibitem{ral}J. Ralston and  P. Jain, Int. J. Mod. Phys. D13, 1857, 2004.
\bibitem{erik1} Eriksen H.K. et al., 2004, Astrophys. J, 605, 14
\bibitem{erik2} Eriksen H.K. et al., 2004, astro-ph/0401276;
astro-ph/0407271
\bibitem{erik3}H. Eriksen et al, ApJ, 612, 633, 2004 [astro-ph/0403098].
\bibitem{hbg}Hansen F.K., Banday  A.J., G\'orski K.M.,2004,astro-ph/0404206
\bibitem{us} Land K., Magueijo J., 2004, astro-ph/0405519 
\bibitem{hansen} Hansen F.K. et al., 2004, astro-ph/0402396
\bibitem{sour}A.Hajian, T.Souradeep, Astrophys.J. 597 (2003).
\bibitem{vielva} Vielva P. et al., 2003, astro-ph/0310273
\bibitem{evil}K. Land and J. Magueijo, astro-ph/0502237.
\bibitem{riaz}J Weeks et al, Mon. Not. Roy. Astron. Soc. 352, 258, 2004.
\bibitem{dodec}B. Roukema et al, astro-ph/0402608.
\bibitem{sperg}N. Cornish et al, Phys. Rev. Lett. 92, 201302, 2004.
\bibitem{sourtop}A. Hajian, T. Souradeep, N. Cornish, Astrophys.J. 618 
(2005) L63-L66.
\bibitem{arj}A. Berera, R. Buniy and T. Kephart, hep-th/0311223.
\bibitem{tesse}T. Jaffe et al, astro-ph/0503213.
\bibitem{moffatinh}J. Moffat, astro-ph/0502110.
\bibitem{bridle}S. Bridle et al,  Mon.Not.Roy.Astron.Soc. 342: L72, 2003.
\bibitem{contaldi}C. Contaldi et al, JCAP 0307: 002, 2003.
\bibitem{slosar}A. Slosar and U. Seljak, Phys. Rev. D70: 083002, 2004.
\bibitem{conf}J. Magueijo, Phys. Lett, B342, 32, 1995.
\bibitem{multipole}K. Land and J. Magueijo, astro-ph/0502574; astro-ph/0407081.
\bibitem{teg0}M. Tegmark, A. de Oliveira-Costa and 
A. Hamilton, Phys.Rev. D68: 123523, 2003.
\bibitem{foregs}C.L. Bennett, et al., 2003, ApJS, 148, 97.
\bibitem{healp} G\'orski K.M., Hivon E., Wandelt B. 1998, astro-ph/9812350
\end{thebibliography}
\end{document}